\documentclass[twocolumn, superscriptaddress]{revtex4-1}
\usepackage{bm}
\usepackage[pdftex]{graphicx}
\usepackage{xcolor}
\usepackage{amsmath}
\usepackage{amssymb}
\usepackage{amsfonts}
\usepackage{amsthm}
\usepackage{url}
\usepackage[utf8]{inputenc}

\usepackage[breaklinks=true,colorlinks=true,linkcolor=blue,urlcolor=blue,citecolor=blue]{hyperref}

\newcommand{\defeq}{\mathrel{\mathop:}=}

\begin{document}

\title{Fluctuating temperature outside superstatistics: thermodynamics of small systems}

\author{Sergio Davis}
\affiliation{Research Center on the Intersection in Plasma Physics, Matter and Complexity, P$^2$MC, Comisión Chilena de Energía Nuclear, Casilla 188-D, Santiago, Chile}
\affiliation{Departamento de F\'isica, Facultad de Ciencias Exactas, Universidad Andres Bello. Sazi\'e 2212, piso 7, Santiago, 8370136, Chile.}

\begin{abstract}
The existence of fluctuations of temperature has been a somewhat controversial topic in thermodynamics but nowadays it is recognized that they must be taken into
account in small, finite systems. Although for nonequilibrium steady states superstatistics is becoming the \emph{de facto} framework for expressing such temperature
fluctuations, some recent results put into question the idea of temperature as a phase space observable. In this work we present and explore the statistics that describes
a part of an isolated system, small enough to have well-defined uncertainties in energy and temperature, but lacking a superstatistical description. These results motivate the use
of the so-called fundamental temperature as an observable and may be relevant for the statistical description of small systems in physical chemistry.
\end{abstract}

\maketitle

\section{Introduction}

Standard thermodynamics successfully describes equilibrium systems consisting of a sufficiently large number of degrees of freedom, either isolated or in
contact with a large enough environment or reservoir. However, finite systems have gained interest because of their potential technological applications, and in them the
fluctuations of quantities such as energy, volume and number of particles cannot be neglected. Remarkably, even when the idea of fluctuations of intensive
quantities such as temperature has generated controversy since several decades ago~\cite{Kittel1973,Kittel1988,Mandelbrot1989} the atomistic simulation community
keeps working within such a paradigm and has brought interesting insights~\cite{Dixit2015,Hickman2016}.

Recently the formalism of superstatistics~\cite{Beck2003}, originally conceived in the context of nonequilibrium steady states of complex, long-range
interacting systems, has been proposed as the foundation for the thermodynamics of small systems, mostly in the context of physical chemistry~\cite{Dixit2013}. However, the identification of the parameter $\beta$ in superstatistics with a fluctuating inverse temperature is
not free of conceptual difficulties~\cite{Davis2018,Sattin2018}.

In this work we present a study of the statistical distributions that describe two parts of a finite, isolated system, providing some general results. Most interesting
is the fact that, although under certain conditions the target system cannot be described by superstatistics, we can nevertheless construct the probability
distribution of the so-called inverse temperature which, we argue, has a physical interpretation and generalizes the temperature of equilibrium, canonical systems.

\section{Temperature in generalized ensembles}

Thermodynamics defines the temperature $T$ using two different routes. On the one hand, in terms of the changes in entropy as
\begin{equation}
\label{eq:thermo_temp}
\frac{1}{T} \defeq \frac{\partial \mathcal{S}}{\partial E},
\end{equation}
and on the other hand, through the parameter \[\beta \defeq 1/k_B T\] appearing in the probability density of microstates $\bm x$ at thermal equilibrium,
\begin{equation}
\label{eq:canon}
P(\bm x|\beta) = \frac{\exp(-\beta H(\bm x))}{Z(\beta)},
\end{equation}
known as the canonical ensemble, where $H(\bm x)$ is the Hamiltonian of the system and $Z(\beta)$ the partition function. We will now extend the definitions in
(\ref{eq:thermo_temp}) and (\ref{eq:canon}) to the general case of a generalized ensemble, i.e. a nonequilibrium steady state of the form
\begin{equation}
\label{eq:rho}
P(\bm x|S) = \rho(H(\bm x))
\end{equation}
for some non-negative function $\rho(E)$, which we will call the \emph{ensemble function}. These generalized ensembles were first studied due to their practical use
in computer simulation of proteins~\cite{Berg2002, Okamoto2004}, but they are in fact also relevant in the study of complex nonequilibrium systems in steady states
using frameworks such as nonextensive statistical mechanics~\cite{Tsallis2009,Naudts2011} and superstatistics.

The definition of temperature in (\ref{eq:thermo_temp}) gives rise to
the \emph{microcanonical inverse temperature} $\beta_\Omega(E)$, a function of the energy defined by
\begin{equation}
\label{eq:microtemp}
\beta_\Omega(E) \defeq \frac{1}{k_B}\frac{\partial \mathcal{S}(E)}{\partial E} = \frac{\partial}{\partial E}\ln \Omega_H(E)
\end{equation}
where $\mathcal{S}(E) \defeq k_B\ln \Omega_H(E)$ is the microcanonical (Boltzmann) entropy and \[\Omega_H(E) \defeq \int d\bm{x}\delta(H(\bm x)-E)\]
is the density of states. The second definition of temperature, as the parameter $\beta$ in the canonical ensemble, gives rise to the
\emph{fundamental inverse temperature} $\beta_F(E)$, which is also a function of the energy and is defined by~\cite{Velazquez2009d, Davis2019}
\begin{equation}
\label{eq:fundtemp}
\beta_F(E; S) \defeq -\frac{\partial}{\partial E}\ln \rho(E; S).
\end{equation}

Using this definition is straightforward to check that, in the particular case of the canonical ensemble, we have \[\beta_F(E; \beta_0) = \beta_0,\] that is, the fundamental
inverse temperature is a constant function.

In a generalized ensemble described by a function $\rho(E; S)$ the probability distribution of the energy is
\begin{equation}
\label{eq:rho_energy}
\begin{split}
P(E|S) & = \int d\bm{x}P(\bm x|S)\delta(H(\bm x)-E) \\
       & = \rho(E)\Omega_H(E),
\end{split}
\end{equation}
and the most probable energy $E^*$ is therefore given by the extremum condition
\begin{equation}
\label{eq:emost}
\frac{\partial}{\partial E}\ln P(E|S)\Big|_{E=E^*} = 0,
\end{equation}
equivalent to the equality of the fundamental and microcanonical (inverse) temperatures, $\beta_F(E^*; S) = \beta_\Omega(E^*)$.

An alternative argument in favor of $\beta_F$ as the natural generalization of inverse temperature in generalized states comes from expanding the entropy
around the most probable energy $E^*$. We have, to first-order, that
\begin{equation}
\label{eq:entropy}
\begin{split}
\mathcal{S} & \defeq -k_B\int d\bm{x} P(\bm x|S)\ln P(\bm x|S) \\
   & = -k_B\big<\ln \rho(E)\big>_S
\end{split}
\end{equation}
can be approximated by \[\mathcal{S} \approx -k_B\ln \rho(E^*) + \beta_F(E^*)\Big[\big<E\big>_S-E^*\Big]\] so the macroscopic definition of
temperature gives
\begin{equation}
\frac{1}{k_B}\left(\frac{\partial \mathcal{S}}{\partial \big<E\big>_S}\right) \approx \beta_F(E^*).
\end{equation}

By making use of the conjugate variables theorem~\cite{Davis2012,Davis2016c} for a probability density $p(x)$ with compact support, namely the identity
\begin{equation}
\left<\frac{\partial \omega(x)}{\partial x}\right>_p + \left<\omega(x)\frac{\partial}{\partial x}\ln p(x)\right>_p = 0,
\end{equation}
where $\omega(x)$ is an arbitrary, differentiable function of $x$, we have for the case of $P(E|S)$ given by (\ref{eq:rho_energy}) that
\begin{equation}
\label{eq:cvt_gen}
\left<\frac{\partial \omega}{\partial E}\right>_S = \Big<\omega\big(\beta_F-\beta_\Omega\big)\Big>_S,
\end{equation}
from which it follows, choosing $\omega = 1$, that
\begin{equation}
\label{eq:equal_temp}
\big<\beta_F\big>_S = \big<\beta_\Omega\big>_S
\end{equation}
that is, the expectation values of both inverse temperatures are equal. We will therefore define the inverse temperature of a generalized ensemble without
ambiguity as
\begin{equation}
\label{eq:betas_def}
\beta_S \defeq \big<\beta_F\big>_S = \big<\beta_\Omega\big>_S.
\end{equation}

\section{Superstatistics}

Among several frameworks and theories developed with the original aim of describing the generalized ensembles useful for nonequilibrium systems such as
plasmas~\cite{Ourabah2015,Davis2019b,Ourabah2020b}, fluids~\cite{Reynolds2003,Gravanis2021}, self-gravitating systems~\cite{Ourabah2020c} and other complex
systems~\cite{Schafer2018,Denys2016}, superstatistics~\cite{Beck2003,Beck2004} is particularly elegant and compact.

The superstatistical framework can be recovered from the single assumption that the inverse temperature $\beta$ in (\ref{eq:canon}) is promoted to a random
variable with a probability density $P(\beta|S)$, and a direct consequence of this assumption is that there is a joint probability distribution function of
the microstate and the inverse temperature, that we can write as
\begin{equation}
\label{eq:joint_superstat}
P(\bm x, \beta|S) = \left[\frac{\exp(-\beta H(\bm x))}{Z(\beta)}\right]P(\beta|S).
\end{equation}

Consequently, if we want to determine the marginal probability density of the microstates, one needs to integrate out the parameter $\beta$, obtaining
\begin{equation}
\label{eq:ensemble_superstat}
\begin{split}
P(\bm x|S) & = \int_0^\infty d\beta P(\bm x, \beta|S) \\
           & = \int_0^\infty d\beta \left[\frac{\exp(-\beta H(\bm x))}{Z(\beta)}\right]P(\beta|S).
\end{split}
\end{equation}

The ensemble in (\ref{eq:ensemble_superstat}) is a generalized ensemble according to (\ref{eq:rho}), with
\begin{equation}
\rho(E; S) = \int_0^\infty d\beta f(\beta)\exp(-\beta E)
\end{equation}
that is, $\rho(E; S)$ is the Laplace transform of a non-negative function $f(\beta) \defeq P(\beta|S)/ Z(\beta)$.

Now we will briefly review some conditions that any superstatistical model must fulfill, using the language of microcanonical and fundamental inverse temperatures. From
the well-known canonical distribution of energies \[P(E|\beta) = \frac{\exp(-\beta E)}{Z(\beta)}\Omega_H(E)\] we can write the joint distribution of energy and inverse
temperature for superstatistics as
\begin{equation}
\label{eq:ebeta}
\begin{split}
P(E, \beta|S) & = P(E|\beta)P(\beta|S) \\
   & = \exp(-\beta E)f(\beta)\Omega_H(E),
\end{split}
\end{equation}
and by again using the conjugate variables theorem, this time with respect to $E$ in $P(E, \beta|S)$, we obtain
\begin{equation}
\label{eq:cvt_ebeta}
\Big<\frac{\partial \omega}{\partial E}\Big>_S = \Big<\omega\big(\beta - \beta_\Omega\big)\Big>_S
\end{equation}
so that the choice $\omega = 1$ gives us
\begin{equation}
\big<\beta\big>_S = \big<\beta_\Omega\big>_S = \beta_S.
\end{equation}

This confirms that our definition of inverse temperature $\beta_S$ agrees with the superstatistical mean inverse temperature. Moreover, using $\omega = \beta_\Omega$
in (\ref{eq:cvt_ebeta}) yields
\begin{equation}
\bigg<\frac{\partial \beta_\Omega}{\partial E}\bigg>_S = \big<\beta\beta_\Omega\big>_S - \Big<(\beta_\Omega)^2\Big>_S
\end{equation}
while using $\omega = \beta$ produces
\begin{equation}
\label{eq:lema1}
\big<\beta^2\big>_S = \big<\beta\beta_\Omega\big>_S.
\end{equation}

By adding and substracting $(\beta_S)^2$ on both sides and using (\ref{eq:lema1}) we have
\begin{equation}
\label{eq:varbeta}
\Big<(\delta \beta)^2\Big>_S = \Big<(\delta \beta_\Omega)^2\Big>_S + \bigg<\frac{\partial \beta_\Omega}{\partial E}\bigg>_S,
\end{equation}
a relation which connects the variance of the superstatistical $\beta$ with the variance of $\beta_\Omega$. Note that the formula (\ref{eq:varbeta}) was
already hinted at in Ref.~\cite{Davis2020} (Eq. (38)) for a particular case of superstatistics, now it is shown to be a general feature of the theory. In
fact, a direct consequence of (\ref{eq:varbeta}) is that an ensemble where
\begin{equation}
\label{eq:U}
\mathcal{U} \defeq \Big<(\delta \beta_\Omega)^2\Big>_S + \Big<\frac{\partial \beta_\Omega}{\partial E}\Big>_S
\end{equation}
is negative cannot be reduced to superstatistics. Higher moments of $\beta$ can be obtained by computing the $n$-th derivative of $\rho(E; S)$ as
\begin{equation}
\frac{1}{\rho}\frac{\partial^n \rho}{\partial E^n} = (-1)^n\int_0^\infty dE \left[\frac{f(\beta)\exp(-\beta E)}{\rho(E; S)}\right]\beta^n
\end{equation}
and using (\ref{eq:ebeta}) together with using Bayes' theorem~\cite{Sivia2006} we recognize the quantity in square brackets as
\begin{equation}
\label{eq:pbeta_cond}
P(\beta|E, S) = \frac{P(E, \beta|S)}{P(E|S)} = \frac{\exp(-\beta E)f(\beta)}{\rho(E; S)},
\end{equation}
so the $n$-th moment of $\beta$ conditional on the value of $E$ is
\begin{equation}
\big<\beta^n\big>_{E,S} = \frac{(-1)^n}{\rho}\frac{\partial^n \rho}{\partial E^n}.
\end{equation}

\noindent
Setting $n = 1$ gives
\begin{equation}
\label{eq:fundbeta_super}
\big<\beta\big>_{E,S} = -\frac{\partial}{\partial E}\ln \rho(E; S) = \beta_F(E)
\end{equation}
and this allows us to interpret the fundamental inverse temperature as the mean superstatistical inverse temperature at fixed $E$. On the other hand, $n = 2$ yields
\begin{equation}
\big<\beta^2\big>_{E,S} = \frac{1}{\rho}\frac{\partial^2 \rho}{\partial E^2}
\end{equation}
and together with (\ref{eq:fundbeta_super}) we can construct the conditional variance of $\beta$ given $E$ as
\begin{equation}
\begin{split}
\Big<(\delta \beta)^2\Big>_{E, S} & = \Big<\beta^2\Big>_{E,S} - \big<\beta\big>^2_{E,S} \\
  & = \frac{1}{\rho}\frac{\partial^2 \rho}{\partial E^2} - \Big(\frac{\partial}{\partial E}\ln \rho\Big)^2 \\
  & = \frac{\partial^2}{\partial E^2}\ln \rho = -\frac{\partial \beta_F(E)}{\partial E}.
\end{split}
\end{equation}

\noindent
Therefore for superstatistics it must hold that
\begin{equation}
\label{eq:super_ineq}
\frac{\partial \beta_F(E)}{\partial E} \leq 0,
\end{equation}
an important result that we will use later on. If we use $\omega = \beta_F$ and $\omega = \beta_\Omega$ in (\ref{eq:cvt_gen}) we obtain
\begin{subequations}
\begin{align}
\label{eq:lema2a}
\Big<\frac{\partial \beta_F}{\partial E}\Big>_S & = \Big<\beta_F^2\Big>_S - \Big<\beta_F\beta_\Omega\Big>_S, \\
\label{eq:lema2b}
\Big<\frac{\partial \beta_\Omega}{\partial E}\Big>_S & = -\Big<\beta_\Omega^2\Big>_S + \Big<\beta_F\beta_\Omega\Big>_S
\end{align}
\end{subequations}
respectively. Adding both and using (\ref{eq:equal_temp}) we can write $\mathcal{U}$ in terms of $\beta_F$ as
\begin{equation}
\label{eq:U2}
\mathcal{U} = \Big<(\delta \beta_F)^2\Big>_S - \Big<\frac{\partial \beta_F}{\partial E}\Big>_S
\end{equation}
which again, due to (\ref{eq:super_ineq}), confirms $\mathcal{U} \geq 0$ for superstatistics, with equality corresponding to the
canonical ensemble. If a system is stable in the canonical ensemble at inverse temperature $\beta$ then, using $\mathcal{U}=0$ in (\ref{eq:U}) we have
\begin{equation}
\Big<\frac{\partial \beta_\Omega}{\partial E}\Big>_\beta = -\Big<(\delta \beta_\Omega)^2\Big>_\beta \leq 0
\end{equation}
which is the condition of positive heat capacity, as we will see in the following sections. For the same system in a general superstatistical state $S$,
because of (\ref{eq:varbeta}) it must hold that $\big<(\delta \beta)^2\big>_S \leq \big<(\delta \beta_\Omega)^2\big>_S$, and because (\ref{eq:U2}) implies
$\big<(\delta \beta)^2\big>_S \geq \big<(\delta \beta_F)^2\big>_S$, it also follows that
\begin{equation}
\label{eq:var_ineq}
\Big<(\delta \beta_\Omega)^2\Big>_S \geq \Big<(\delta \beta_F)^2\Big>_S.
\end{equation}

\section{A composite system in the microcanonical ensemble}
\label{sec:theorem}

\begin{figure}[b!]
\begin{center}
\includegraphics[height=5cm]{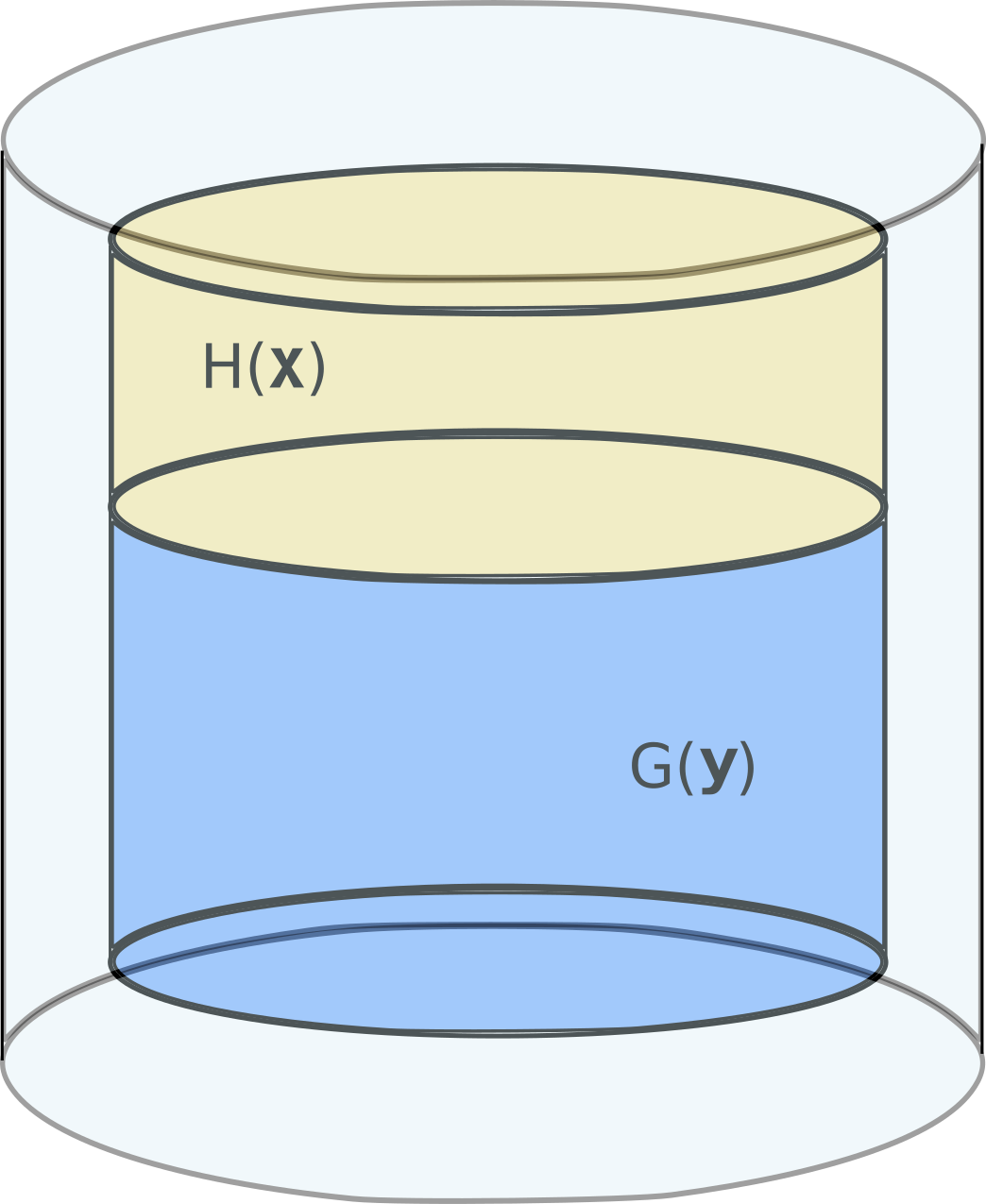}
\end{center}
\caption{Diagram representing two systems in contact with energies $H(\bm x)$ and $G(\bm y)$ so that $H(\bm x)+G(\bm y) = E_0$ remains fixed.}
\label{fig:cylinders}
\end{figure}

\noindent
Consider two systems in contact, forming a \emph{composite system} which is isolated from the rest of the universe. We will focus on one of the systems, the \emph{target}, with
degrees of freedom $\bm x$ while the other will be the \emph{environment} with degrees of freedom $\bm y$, as depicted in Fig.~\ref{fig:cylinders}. The (fixed) energy of the
composite system is given by the sum \[\mathcal{H}(\bm x, \bm y) = H(\bm x) + G(\bm y) = E_0,\]
and the joint probability density of $\bm x$ and $\bm y$ is given by the microcanonical ensemble
\begin{equation}
P(\bm x, \bm y|E_0) = \frac{1}{\Omega_\mathcal{H}(E_0)}\delta(H(\bm x)+G(\bm y)-E_0).
\end{equation}

The marginal probability density for the target system is then obtained~\cite{Kardar2007} by integrating over the environment
\begin{equation}
\label{eq:marginal_x}
\begin{split}
P(\bm x|E_0) & = \frac{1}{\Omega_\mathcal{H}(E_0)}\int d\bm{y}\delta(H(\bm x)+G(\bm y)-E_0) \\
             & = \frac{\Omega_G(E_0-H(\bm x))}{\Omega_\mathcal{H}(E_0)}
\end{split}
\end{equation}
where \[\Omega_G(G) \defeq \int d\bm{y}\delta(G(\bm y)-G)\]
is the density of states of the environment. We see that the target system is described by a generalized ensemble with ensemble function
\begin{equation}
\rho(E; E_0) = \frac{\Omega_G(E_0-E)}{\Omega_\mathcal{H}(E_0)}
\end{equation}
and fundamental inverse temperature given by
\begin{equation}
\label{eq:betaf_target}
\beta_F^{(x)}(E; E_0) = -\frac{\partial}{\partial E}\ln \Omega_G(E_0-E) = \beta_\Omega^{(y)}(E_0-E).
\end{equation}

Here we see that the functional form of the fundamental inverse temperature, and therefore of the ensemble function $\rho$ itself, is completely determined
by the functional form of the microcanonical inverse temperature of the environment. Please note also that, because
\begin{equation*}
\begin{split}
\Big<\beta_F^{(x)}\Big>_S & = \int dE P(E|S) \beta_F^{(x)}(E; E_0) \\
  & = \int dE P(E|S)\beta_\Omega^{(y)}(E_0-E) \\
  & = \int dG P(G|S)\beta_\Omega^{(y)}(G) = \Big<\beta_\Omega^{(y)}\Big>_S
\end{split}
\end{equation*}
and using (\ref{eq:equal_temp}) and (\ref{eq:betas_def}) it must hold that $\beta_S^{(x)} = \beta_S^{(y)}$.

\noindent
Taking the derivative of (\ref{eq:betaf_target}) with respect to $E$ we have
\begin{equation}
\label{eq:main}
\frac{\partial \beta_F^{(x)}}{\partial E} = -\frac{\partial \beta_\Omega^{(y)}}{\partial G}\Big|_{G=E_0-E},
\end{equation}
and using the relation~\cite{Velazquez2009a}
\begin{equation}
\label{eq:thermo_cv}
\frac{\partial \beta(E)}{\partial E} = \frac{1}{k_B}\frac{\partial}{\partial E}\left(\frac{1}{T(E)}\right) = -\frac{\beta(E)^2}{C_V(E)}
\end{equation}
we can write
\begin{equation}
\label{eq:condition}
\frac{\partial \beta_F^{(x)}}{\partial E} = \frac{\big(\beta_\Omega^{(y)}\big)^2}{C_V^{(y)}}
\end{equation}
where $C_V=(\partial E/\partial T)_V$ is the microcanonical heat capacity in units of $k_B$. Because superstatistics requires (\ref{eq:super_ineq})
for all energies, it follows from (\ref{eq:condition}) that a target system cannot be described by superstatistics if it is enclosed together with an environment
with positive heat capacity, forming an isolated system of target plus environment.

\section{The microcanonical distribution of configurations}

A simple but illustrative example of this last result ocurrs for the microcanonical distribution of configurations~\cite{Severin1978,Ray1991}
\begin{equation}
\label{eq:ray}
P(\bm{R}|E_0)  = \frac{\omega_0}{\Omega_\mathcal{H}(E_0)}\Big[E_0 - \Phi(\bm{R})\Big]_+^{\frac{3N}{2}-1}
\end{equation}
for a system of classical particles with Hamiltonian \[\mathcal{H}(\bm{R}, \bm{P}) = K(\bm{P}) + \Phi(\bm{R})\]
kept at total energy $E_0$, where $K(\bm{P})$ is the non-relativistic kinetic energy \[K(\bm{P}) = \sum_{i=1}^N \frac{\bm{p}_i^2}{2 m_i}\]
and $\Phi(\bm{R})$ is the interaction energy.

This is a particular case of our result of Section \ref{sec:theorem}, given that we can interpret (\ref{eq:ray}) as the distribution of a ``purely configurational'' target
system placed in contact with an ``ideal gas'' environment with density of states $\Omega_K(k) = \omega_0 k^{\frac{3N}{2}-1}$ in such a manner that the composite
system of target plus environment is isolated.

As noted by Naudts and Baeten~\cite{Naudts2009}, the result in (\ref{eq:ray}) is a case of the Tsallis $q$-exponential distribution
\begin{equation}
\label{eq:qexpon}
P(\bm x|\beta_0, q) \propto \Big[1+(q-1)\beta_0 H(\bm x)\Big]_+^{\frac{1}{1-q}}
\end{equation}
with \emph{entropic index} $q = 1 - \tfrac{2}{3N-2} \leq 1$. The fundamental inverse temperature corresponding to (\ref{eq:qexpon}) is~\cite{Davis2019,Umpierrez2021}
\begin{equation}
\beta_F(E; \beta_0, q) = \frac{\beta_0}{1+(q-1)\beta_0 E},
\end{equation}
with derivative \[\frac{\partial \beta_F(E; \beta_0, q)}{\partial E} = (1-q)\beta_F(E; \beta_0, E)^2\] which is positive for $q < 1$, hence $P(\bm R|E_0)$ in
(\ref{eq:ray}) cannot be described by superstatistics, unless $q = 1$ which is the canonical ensemble, situation that only ocurrs in the thermodynamic limit
$N \rightarrow \infty$. That is, no function $f(\beta)$ exists
such that
\begin{equation}
\int_0^\infty\hspace{-5pt} d\beta f(\beta)\exp(-\beta \Phi(\bm R)) = \frac{\omega_0}{\Omega_\mathcal{H}}\Big[E_0-\Phi(\bm R)\Big]_+^{\frac{3N}{2}-1}
\end{equation}
for finite $N$, and in fact, the inverse Laplace transform of the right-hand side would yield
\begin{equation}
f(\beta) = -\frac{\omega_0\exp(\beta E_0)(-\beta)^{1-\frac{3N}{2}}}{\Omega_\mathcal{H}(E_0)\Gamma(1-\tfrac{3N}{2})}
\end{equation}
but this is undefined for $N \geq 1$ because of the $\Gamma$-function with negative argument. We might be tempted to conclude that there is no fluctuating
temperature in the microcanonical ensemble, however we still have the definitions of $\beta_\Omega$ in (\ref{eq:microtemp}) and $\beta_F$ in (\ref{eq:fundtemp}).
In particular, for $\bm R$ the later is given by
\begin{equation}
\beta_F(\bm R) = \frac{3N-2}{2(E_0-\Phi(\bm R))}
\end{equation}
with a clear interpretation related to the equipartition theorem,
\begin{equation}
E_0 = \Phi(\bm R) + \Big(\tfrac{3N-2}{2}\Big)k_B T_F(\bm R)
\end{equation}
where $k_B T_F = 1/\beta_F$. This suggests considering the instantaneous value of the (inverse) fundamental temperature as a consistent definition
of fluctuating temperature outside superstatistics.

\section{Some results for a fixed ratio between target and environment}

In this section we will consider a setup of target and environment with $N + N_e$ particles, $N$ forming the target and $N_e$ the environment, with a fixed
ratio $\gamma \defeq N/N_e$. This condition allows us to take the thermodynamic limit $N_e(1+\gamma) \rightarrow \infty$ for the composite system without making
the environment infinitely larger than the target, thus avoiding the canonical ensemble.

We will assume an environment $\bm y$ with energy $G(\bm y)$ undergoing small fluctuations around a value $G_0$. We can then approximate the logarithm of the
density of states to second order as
\begin{equation}
\ln \Omega_G(G) \approx \ln \Omega_G(G_0) + \beta_e(G-G_0) - \frac{B}{2}(G-G_0)^2
\end{equation}
with
\begin{subequations}
\begin{align}
\beta_e & \defeq \frac{\partial}{\partial G_0}\ln \Omega_G(G_0), \\
B & \defeq -\frac{\partial^2}{\partial {G_0}^2}\ln \Omega_G(G_0).
\end{align}
\end{subequations}

Note that $\beta_e$ is the microcanonical inverse temperature of the environment at $G_0$ and $B$ is connected to the heat capacity $C$ at $G_0$ according to
(\ref{eq:thermo_cv}) by $C \defeq (\beta_e)^2 / B$. After completing the square we have
\begin{equation}
\label{eq:env_dos}
\Omega_G(G) = \omega_1\exp\left(-A\Big[1-\frac{G}{G_\text{max}}\Big]^2\right)
\end{equation}
where we have defined
\begin{subequations}
\begin{align}
G_\text{max} & \defeq G_0 + \frac{C}{\beta_e}, \\
A & \defeq \frac{(\beta_e G_\text{max})^2}{2C}, \\
\ln \omega_1 & \defeq \ln \Omega_G(G_0) + \frac{C}{2}
\end{align}
\end{subequations}
with $C > 0$. With these choices and defining $g \defeq G/G_\text{max}$ the density of states is monotonically increasing because
\begin{equation}
\label{eq:beta_env}
\beta_\Omega^{(y)}(G) = \frac{2A}{(G_\text{max})^2}(G_\text{max}-G) \geq 0
\end{equation}
and concave because
\begin{equation}
\label{eq:environ_neg}
\frac{\partial \beta_\Omega^{(y)}}{\partial G} = -\frac{(\beta_e)^2}{C} \leq 0,
\end{equation}
as expected of a well-behaved thermodynamic system away from a phase transition. We must also ensure that the approximation preserves the extensivity property
of the entropy. This imposes that \[\lim_{N_e \rightarrow \infty} \frac{\mathcal{S}}{N_e} = s\] with $s$ independent of $N$~\cite{Touchette2002}. Given that $\beta_e$ is
intensive and both $C$ and $G_\text{max}$ are extensive, it follows that $A$ is extensive and we define
\begin{equation}
a \defeq \frac{A}{N_e} = \frac{(\beta_e \gamma \varepsilon_0)^2}{2c'}
\end{equation}
where $c' \defeq C/N_e$. The heat capacity of the environment is given by (\ref{eq:thermo_cv}) as \[\frac{1}{C_V^{(y)}} = -\frac{1}{(\beta_\Omega^{(y)})^2}\frac{\partial \beta_\Omega^{(y)}}{\partial G}\] therefore
\begin{equation}
C_V^{(y)} = 2 a N_e(1-g)^2 \geq 0,
\end{equation}
which is extensive, as expected. The entropy per particle is
\begin{equation}
\begin{split}
\frac{\mathcal{S}_2}{N_e k_B} & = \frac{1}{N_e}\ln \Omega_G(G) \\
  & \approx \frac{1}{N_e}\ln \Omega_G(G_0) + \frac{c'}{2} - a(1-g)^2
\end{split}
\end{equation}
thus for it to be extensive it is required that
\begin{equation}
\label{eq:sg}
s_G \defeq \lim_{N_e \rightarrow \infty} \frac{1}{N_e}\ln \Omega_G(G_0)
\end{equation}
is finite. Thus we see that this approximation to the density of states of the environment preserves its properties relevant to stability and extensivity.

\begin{figure}
\begin{center}
\includegraphics[width=8cm]{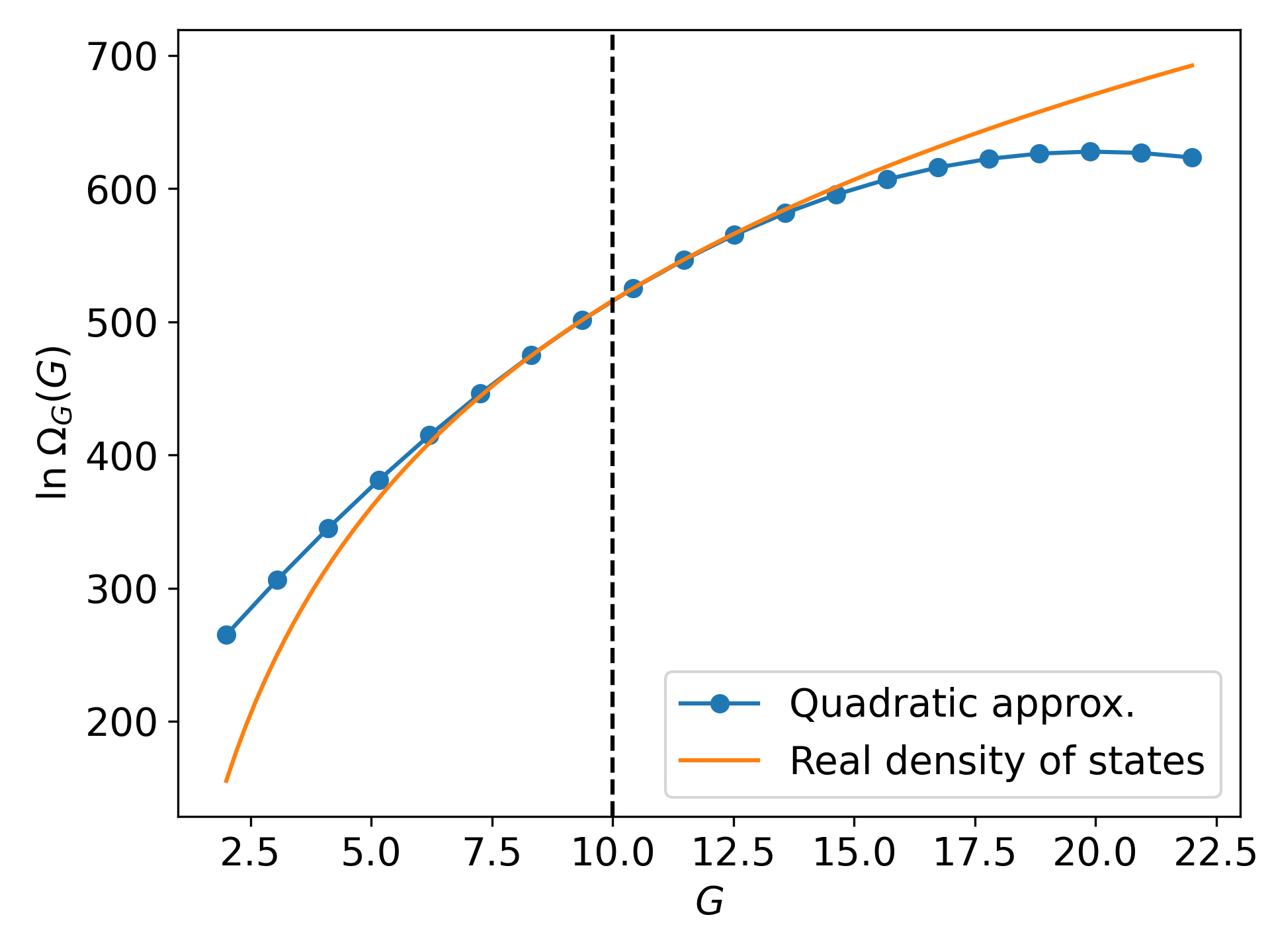}
\end{center}
\caption{Logarithm of the density of states of an ideal gas of $N_e = 150$ particles, $c' = 3/2$ and $G_0$ = 10. The solid orange line is the exact density
of states while the blue dots represent the quadratic approximation of (\ref{eq:env_dos}).}
\end{figure}

However, it follows from (\ref{eq:main}), (\ref{eq:environ_neg}) and (\ref{eq:super_ineq}) that the resulting target distribution $P(\bm x|E_0)$ induced by contact with
this environment cannot be described by superstatistics. We will now explore the features of this induced ensemble.

If we set the minimum value of $H(\bm x)$ as zero, we are forced to set $G_\text{max} = E_0$ and we obtain the induced ensemble for the target system as
\begin{equation}
\label{eq:model}
\begin{split}
P(\bm x|S) & = \frac{\Omega_G(E_0-E)}{\Omega_\mathcal{H}(E_0)} \\
           & = \frac{\omega_1}{\Omega_\mathcal{H}(E_0)}\exp\left(-A\big(\tfrac{H(\bm x)}{E_0}\big)^2\right)
\end{split}
\end{equation}
with $0 \leq H(\bm x) \leq E_0$ and where $S \defeq (E_0, A)$ is a shortcut that encodes the ensemble parameters. Defining the normalization constant
\begin{equation}
\mathcal{Z} \defeq \frac{\Omega_\mathcal{H}(E_0)}{\omega_1} = \int d\bm{x}\exp(-A(H(\bm x)/E_0)^2)
\end{equation}
we can write the ensemble function associated to (\ref{eq:model}) as
\begin{equation}
\label{eq:rho_target}
\rho(E; S) = \frac{\exp(-A(E/E_0)^2)}{\mathcal{Z}}.
\end{equation}
which is a particular case of the Gaussian ensemble~\cite{Challa1988a, Johal2003}. This is almost the same expansion as discussed by Ramshaw~\cite{Ramshaw2018}, except in our
case we fix beforehand the ratio $\gamma$ between sizes of target and environment, thus (\ref{eq:rho_target}) can no longer be reduced to the canonical ensemble.

In order to compute the thermodynamic properties of the target system, we need knowledge of its Hamiltonian, or equivalently, its density of states. For that purpose we
will consider the target system as composed of $N$ particles, with with density of states of the form
\begin{equation}
\begin{split}
\Omega_H(E) & = \omega(N) E^{cN-1} \\
\text{with}\quad \omega(N) & \defeq \frac{(N \zeta)^N}{N!\Gamma(cN)}
\end{split}
\end{equation}
and where $c > 0$ and $\zeta > 0$ are constants, independent of $N$. The partition function is
\begin{equation}
\begin{split}
Z(\beta) & = \omega(N)\int_0^\infty dE\exp(-\beta E)E^{cN-1} \\
         & = \frac{(N\zeta\beta^{-c})^N}{N!}
\end{split}
\end{equation}
which gives a canonical caloric curve
\begin{equation}
\big<E\big>_\beta = -\frac{\partial}{\partial \beta}\ln Z(\beta) = \frac{cN}{\beta}
\end{equation}
hence this choice of $\Omega_H(E)$ encompasses all systems with constant, positive heat capacity $C_V^{(x)} = cN k_B$. The microcanonical entropy is extensive,
and its value per particle is equal to
\begin{equation}
\label{eq:micro_entropy}
\begin{split}
\frac{\mathcal{S}(E)}{N k_B} & \defeq \lim_{N \rightarrow \infty}\frac{1}{N}\ln \Omega_H(E) \\
   & = 1+\ln \zeta + c\left(1+\ln \frac{\varepsilon}{c}\right)
\end{split}
\end{equation}
where $\varepsilon \defeq E/N$ is the energy per particle. The energy distribution is readily obtained by (\ref{eq:rho_energy}) which in this case gives
\begin{equation}
\label{eq:energy_dist}
P(E|S) = \frac{\omega(N)}{\mathcal{Z}}\exp(-A(E/E_0)^2)E^{cN-1}
\end{equation}
with
\begin{equation}
\label{eq:partition_trunc}
\begin{split}
\mathcal{Z} & = \omega(N)\int_0^{E_0} dE\exp(-A(E/E_0)^2)E^{cN-1} \\
            & = \frac{\omega(N)}{2}\Big(\frac{E_0}{\sqrt{A}}\Big)^{cN}\Gamma(\tfrac{cN}{2}; 0 \rightarrow A)
\end{split}
\end{equation}
where \[\Gamma(z; a \rightarrow b) \defeq \int_a^b dt \exp(-t)t^{z-1} \] is the incomplete $\Gamma$-function. Here we will make the simplifying assumption that
$P(E = E_0|S) \approx 0$ in order to take $E \in [0, \infty)$, and this can always be achieved for large enough system size. In fact,
\[P(E=E_0|S) = \frac{2\exp(-\frac{aN}{\gamma})\sqrt{\frac{aN}{\gamma}}^{cN}}{N\varepsilon_0\Gamma(\tfrac{cN}{2}; 0 \rightarrow \tfrac{aN}{\gamma})} \longrightarrow 0\]
as $N \rightarrow \infty$. Therefore, for large enough $N$ we can replace (\ref{eq:partition_trunc}) by
\begin{equation}
\label{eq:partition}
\begin{split}
\mathcal{Z} & = \omega(N)\int_0^\infty dE \exp(-A(E/E_0)^2)E^{cN-1} \\
            & = \frac{\omega(N)}{2}\Big(\frac{E_0}{\sqrt{A}}\Big)^{cN}\Gamma(\tfrac{cN}{2})
\end{split}
\end{equation}
and replacing (\ref{eq:partition}) into (\ref{eq:energy_dist}) we can write the full, properly normalized energy distribution as
\begin{equation}
\label{eq:energydist0}
P(E|S) = \frac{2(\sqrt{A})^{cN}}{E_0\Gamma(\tfrac{cN}{2})}\exp\left(-A\big(\tfrac{E}{E_0}\big)^2\right)\big(\tfrac{E}{E_0}\big)^{cN-1}.
\end{equation}

The most probable energy $E^*$ is given by the extremum condition in (\ref{eq:emost}), and in this case we obtain
\begin{equation}
\label{eq:estar}
E^* = E_0\sqrt{\frac{c\gamma}{2a}}.
\end{equation}

Now, because $E^* \leq E_0$, we must have $\gamma \leq 2a/c$. We can write the probability density of the energy per particle only in terms of $N$ and
intensive quantities, as
\begin{equation}
\label{eq:energy_dist_per}
P(\varepsilon|S) = \frac{2\beta_e\exp\left(-\tfrac{N\gamma}{2c'}(\beta_e\varepsilon)^2\right)(\beta_e \varepsilon)^{cN-1}}{\left(\tfrac{2c'}{N\gamma}\right)^{\frac{cN}{2}}\Gamma(\tfrac{cN}{2})}
\end{equation}
with most probable value
\begin{equation}
\label{eq:eperstar}
\varepsilon^* = \frac{1}{\beta_e}\sqrt{\frac{c'(cN-1)}{N\gamma}} \approx \frac{1}{\beta_e}\sqrt{\frac{c\:c'}{\gamma}}.
\end{equation}

\begin{figure}
\begin{center}
\includegraphics[width=8cm]{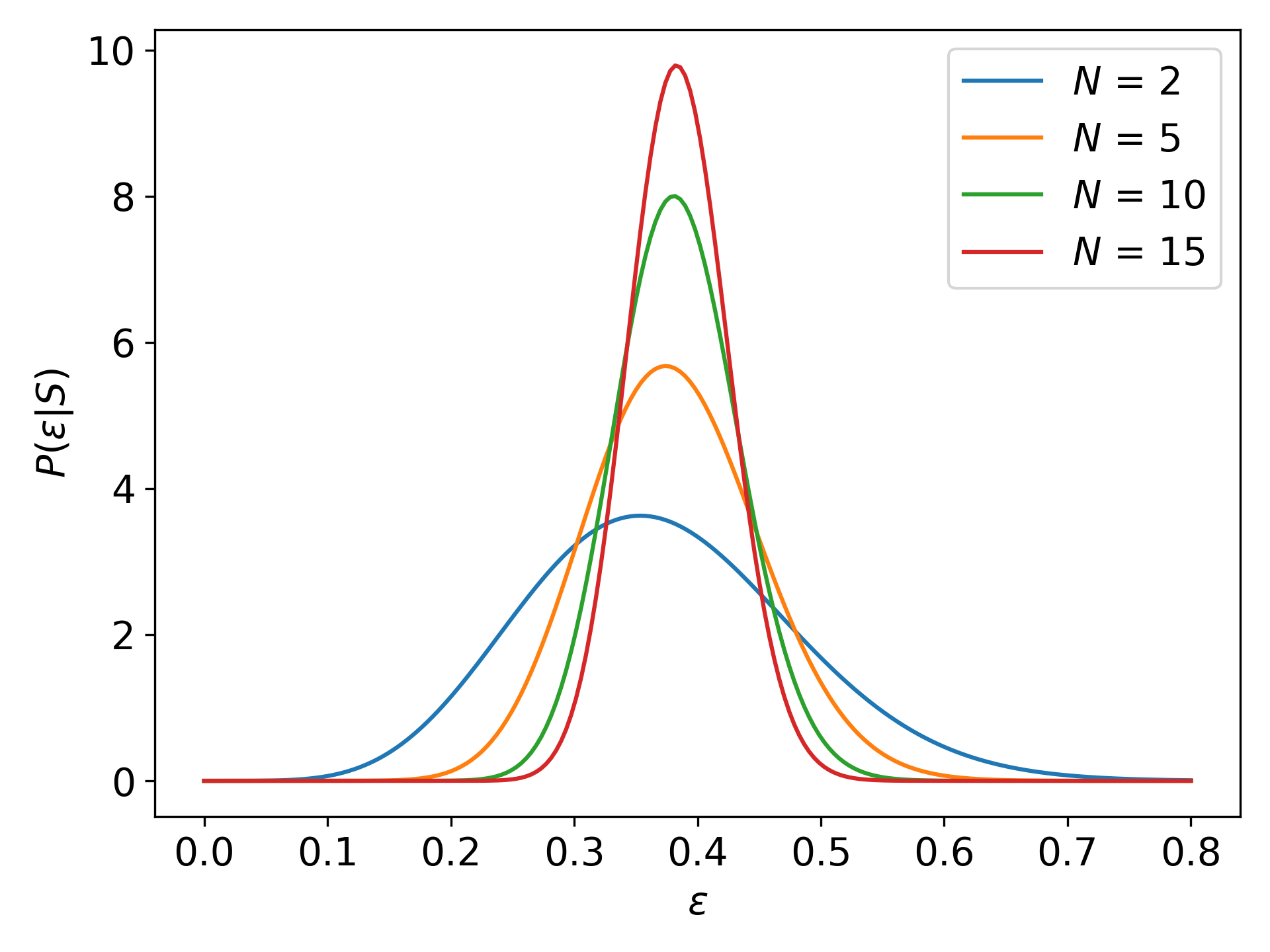}
\end{center}
\caption{Probability density of energy per particle, as given by (\ref{eq:energy_dist_per}), for different sizes of the target system with $\varepsilon_0=1$, $c = 3$, $a = 1$
and $\gamma = 1/10$. We can see that even for small system sizes the probability of having $\varepsilon = \varepsilon_0$ is negligible.}
\end{figure}

In the Gaussian approximation for large enough $N$, the distribution in (\ref{eq:energy_dist_per}) has a variance equal to
\begin{equation}
\label{eq:var}
(\Delta \varepsilon)^2 = \frac{(\varepsilon^*)^2}{2(cN-1)} \approx \frac{(\varepsilon^*)^2}{2cN}
\end{equation}
so that $\Delta \varepsilon \rightarrow 0$ when $N \rightarrow \infty$ and in this way we have
\begin{equation}
\label{eq:delta_dist}
\lim_{N \rightarrow \infty} P(\varepsilon|S) = \delta(\varepsilon-\varepsilon^*).
\end{equation}

\noindent
It is convenient to define a scale of inverse temperature through the constant
\begin{equation}
\beta_0 \defeq \frac{\sqrt{A}}{E_0} = \beta_e\sqrt{\frac{\gamma}{2c'N}}
\end{equation}
which allows us to write (\ref{eq:energydist0}) as
\begin{equation}
\label{eq:energydist}
P(E|S) = \frac{2\beta_0}{\Gamma(\tfrac{cN}{2})}\exp(-(\beta_0 E)^2)(\beta_0 E)^{cN-1},
\end{equation}
noting that $1/\beta_0$ acts as a scale parameter for the energy. The mean value of the energy is
\begin{equation}
\label{eq:emean}
\big<E\big>_S = \int_0^\infty dE P(E|S)E = \frac{1}{\beta_0}\frac{\Gamma\big(\tfrac{cN+1}{2}\big)}{\Gamma\big(\tfrac{cN}{2}\big)},
\end{equation}
and instead of computing the rest of the moments of $E$ from the integrals using $P(E|S)$, we will use the technique recently developed in Ref.~\cite{Umpierrez2021}.
For this, we first take the conjugate variables theorem (\ref{eq:cvt_gen}) and replace $\beta_F$ and $\beta_\Omega$, obtaining
\begin{equation}
\Big<\frac{\partial \omega}{\partial E}\Big>_S = \left<\omega\Big(2(\beta_0)^2 E - \tfrac{cN-1}{E}\Big)\right>_S,
\end{equation}
where using the choice $\omega(E) = E^{n+1}$ we can write the recurrence relation
\begin{equation}
\big<E^{n+2}\big>_S = \frac{(n+cN)}{2(\beta_0)^2}\big<E^n\big>_S
\end{equation}
for the moments of $E$. Using $n = 0$, $n = -2$ and $n = -1$ we can readily obtain
\begin{subequations}
\begin{align}
\big<E^2\big>_S & = \frac{cN}{2}(\beta_0)^{-2}, \\
\big<E^{-2}\big>_S & = \frac{2(\beta_0)^2}{cN-2}, \\
\label{eq:plusminus}
\big<E^{-1}\big>_S & = \big<E\big>_S\cdot \frac{2(\beta_0)^{2}}{cN-1}.
\end{align}
\end{subequations}

\noindent
From (\ref{eq:plusminus}) and (\ref{eq:emean}) we directly read
\begin{equation}
\big<E^{-1}\big>_S = \frac{2\beta_0}{cN-1}\frac{\Gamma\big(\tfrac{cN+1}{2}\big)}{\Gamma\big(\tfrac{cN}{2}\big)}
= \beta_0\frac{\Gamma\big(\tfrac{cN-1}{2}\big)}{\Gamma\big(\tfrac{cN}{2}\big)},
\end{equation}
and we now have all that is required to determine the mean and variance of energy and inverse temperature.

\noindent
A general expression for the $n$-th moment is
\begin{equation}
\label{eq:moments}
\big<E^n\big>_S = (\beta_0)^{-n}\left(\frac{g_n}{g_0}\right).
\end{equation}
where we have defined for convenience the set of functions
\begin{equation}
g_n \defeq \Gamma\big(\tfrac{cN+n}{2}\big).
\end{equation}
which have the property
\begin{equation}
\label{eq:jump}
g_{n+2} = g_n\cdot \frac{cN+n}{2}.
\end{equation}

\begin{figure}
\begin{center}
\includegraphics[width=8cm]{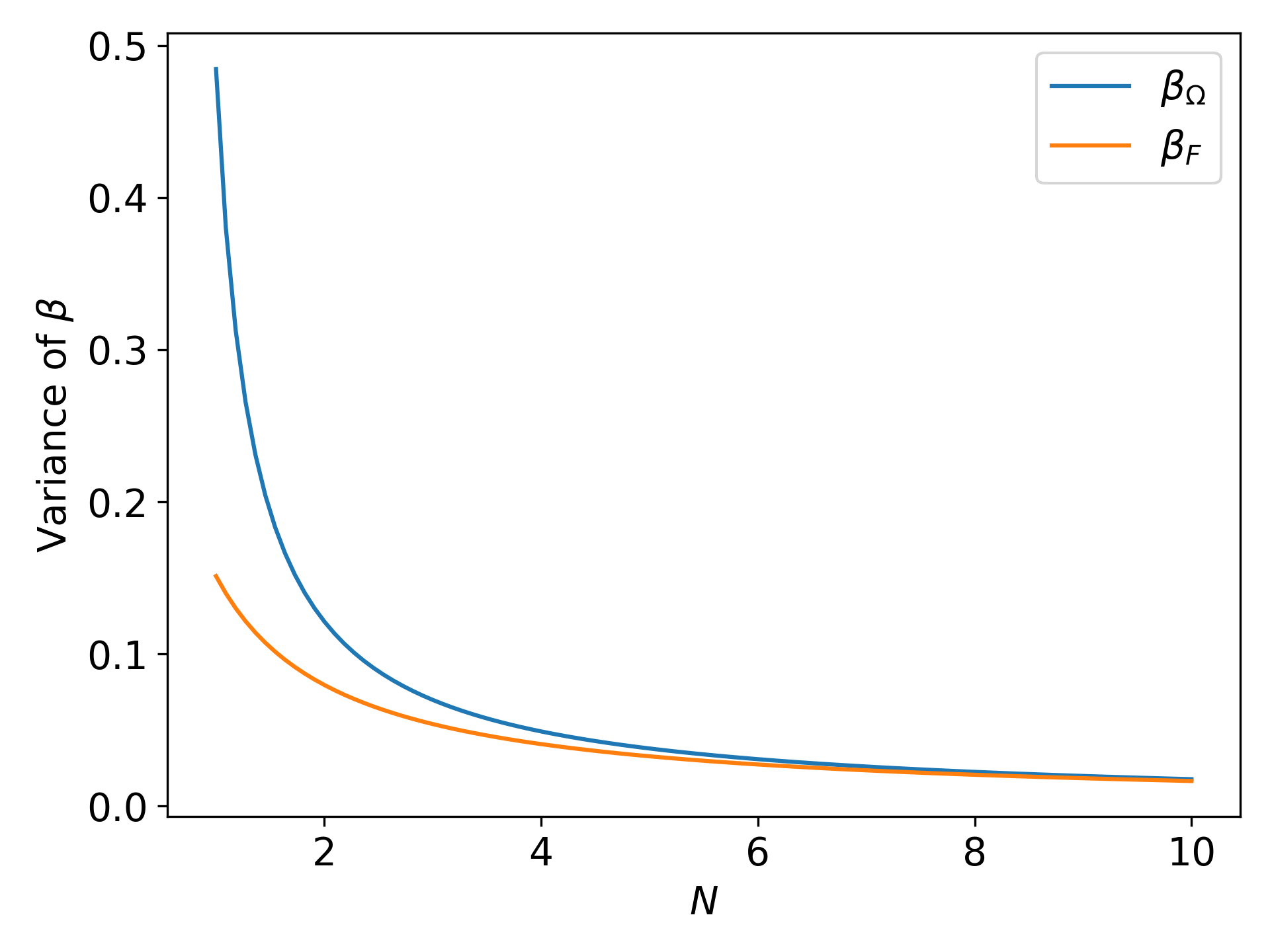}
\end{center}
\caption{Variance of the fundamental and microcanonical inverse temperatures, in units of $\beta^*$, as a function of target system size.}
\label{fig:relvar}
\end{figure}

In order for (\ref{eq:delta_dist}) and (\ref{eq:moments}) to be consistent, it must hold that, for $N \rightarrow \infty$,
\begin{equation}
\label{eq:gratio}
\left(\frac{g_n}{g_0}\right) \approx (N \varepsilon^*\beta_0)^n = \left(\sqrt{\frac{cN}{2}}\right)^n
\end{equation}
which can also be proven by use of the Stirling approximation for $g_n$. Finally, we can write the energy and its relative variance as
\begin{subequations}
\begin{align}
\label{eq:emean2}
\big<E\big>_S & = \frac{1}{\beta_0}\frac{g_1}{g_0}, \\
\label{eq:relvar}
\frac{\big<(\delta E)^2\big>_S}{\big<E\big>_S^2} & = \frac{g_0\cdot g_2}{(g_1)^2} - 1,
\end{align}
\end{subequations}
noting that \[\frac{\big<(\delta E)^2\big>_S}{\big<E\big>_S^2} \rightarrow 0\] as $N \rightarrow \infty$. The fundamental inverse temperature
can be written in terms of $\beta_0$ as
\begin{equation}
\label{eq:betaf}
\beta_F(E; S) = 2(\beta_0)^2 E = \frac{\gamma(\beta_e)^2}{c'}\varepsilon
\end{equation}
and we can verify that it matches $\beta_\Omega^{(y)}(E_0-E)$ according to (\ref{eq:beta_env}), while the microcanonical inverse temperature is equal to
\[\beta_\Omega(E) = \frac{cN-1}{E} \approx \frac{c}{\varepsilon}.\]

We can obtain the inverse temperature $\beta_S$ of the target system through the expectation of either $\beta_\Omega$ or $\beta_F$. The former yields
\begin{equation}
\label{eq:betas}
\big<\beta_\Omega\big>_S = \left<\frac{cN-1}{E}\right>_S
        = (cN-1)\beta_0\Big(\frac{g_{-1}}{g_0}\Big)
\end{equation}
while the later is given by
\begin{equation}
\big<\beta_F\big>_S = 2(\beta_0)^2\big<E\big>_S = (cN-1)\beta_0\Big(\frac{g_{-1}}{g_0}\Big)
\end{equation}
where in the last equality we have used (\ref{eq:emean2}) and (\ref{eq:jump}), confirming the equality in (\ref{eq:equal_temp}). By using (\ref{eq:gratio}) we can approximate
$\beta_S \approx \beta^* $ in the limit $N \rightarrow \infty$, where
\begin{equation}
\label{eq:bstar}
\beta^* \defeq \beta_0 \sqrt{2cN} = \beta_e\sqrt{\frac{c\gamma}{c'}}.
\end{equation}

Here it is interesting to note that, in the case where the target and environment have the same size and heat capacity per particle, that is, $\gamma = 1$ and $c = c'$,
we have $\beta^* = \beta_e$, and because of (\ref{eq:eperstar}), it also holds that $\beta^*\varepsilon^* = c$.

The variance of the fundamental inverse temperature is given by
\begin{equation}
\big<(\delta \beta_F)^2\big>_S = \frac{2(\beta^*)^2}{cN}\left[\frac{cN}{2}- \left(\frac{g_1}{g_{0}}\right)^2\right]
\end{equation}
which goes to zero when $N \rightarrow \infty$, as expected. On the other hand, the variance of the microcanonical inverse temperature is
\begin{equation}
\Big<(\delta \beta_\Omega)^2\Big>_S = \frac{(cN-1)^2(\beta^*)^2}{2cN}\left[\frac{2}{cN-2} - \left(\frac{g_{-1}}{g_0}\right)^2\right].
\end{equation}

We see that both relative variances become closer to each other in the limit $N \rightarrow \infty$, as is shown in Fig.~\ref{fig:relvar}, and we can also note that the
inequality (\ref{eq:var_ineq}) still holds in this case, even when we are not dealing with a superstatistical system.

\begin{figure}
\begin{center}
\includegraphics[width=8cm]{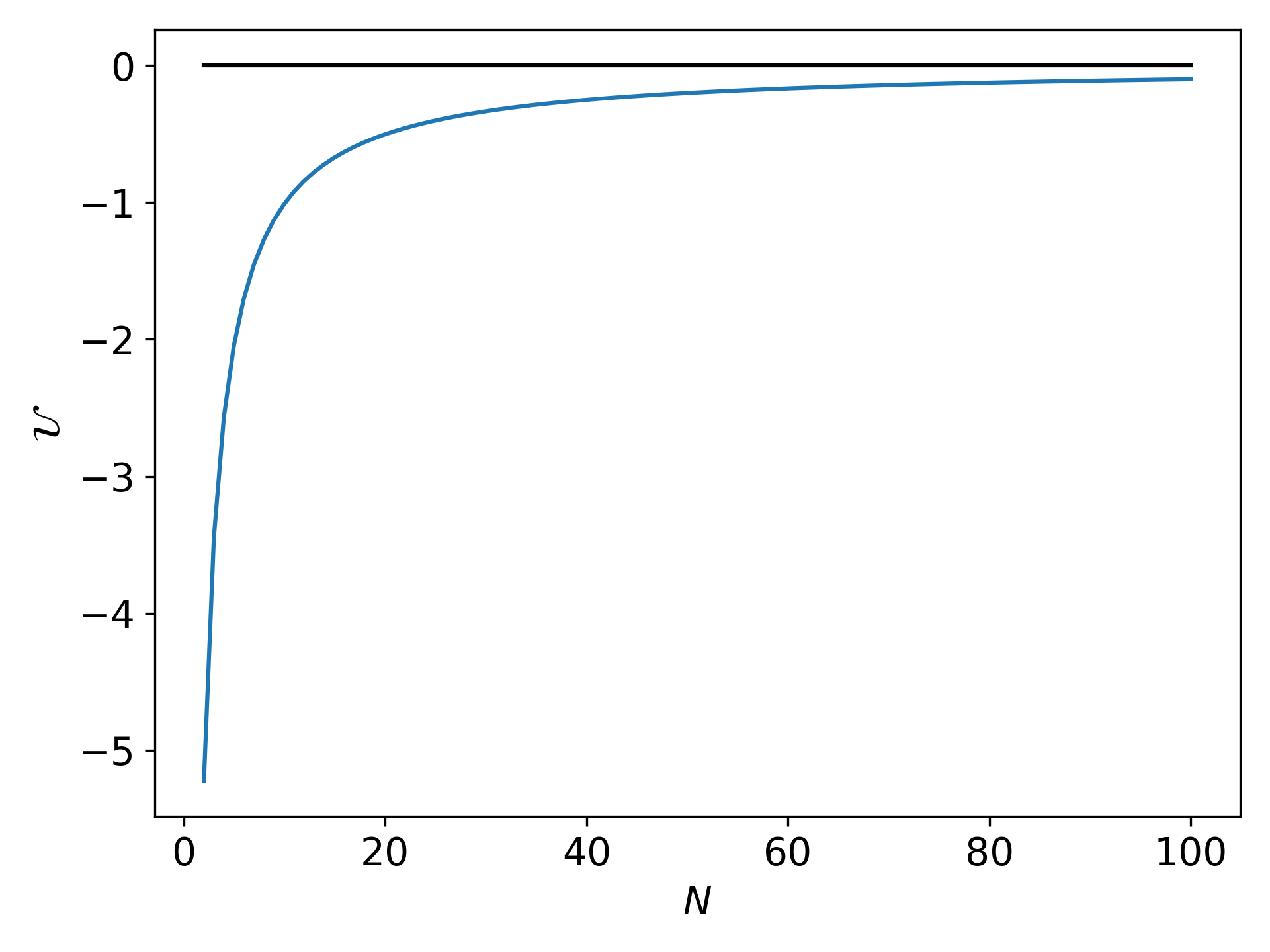}
\end{center}
\caption{The observable $\mathcal{U}$ defined by (\ref{eq:U}) as a function of target system size $N$. This quantity is never positive, as opposed to what is expected in superstatistics.}
\label{fig:quant}
\end{figure}

Another test that reveals the deviation from superstatistics is the sign of the observable $\mathcal{U}$ in (\ref{eq:U}). In our case we have
\[\Big<\frac{\partial \beta_\Omega}{\partial E}\Big>_S = -\left<\frac{cN-1}{E^2}\right>_S = -(cN-1)\beta_0\Big(\frac{g_{-2}}{g_0}\Big)\]
hence
\begin{equation}
\mathcal{U} = ([cN-1]\beta_0)^2\left[\frac{2}{cN-1}-\left(\frac{g_{-1}}{g_0}\right)^2\right]
\end{equation}
which is never positive and goes to zero as $N \rightarrow \infty$, as seen in Fig.~\ref{fig:quant}.

Given that we have the probability distribution of energy, obtaining the corresponding distribution of fundamental and microcanonical inverse
temperature is relatively straightforward. We need only to use the transformation rule \[P(f = F|I) = \int dx P(x|I)\delta(f(x)-F)\] for
probability densities, so that for $\beta_F$ we obtain
\begin{equation}
\label{eq:pbetaf}
P(\beta_F = \beta|S) = \frac{\exp\big(-\frac{1}{4}(\beta/\beta_0)^2\big)(\beta/\beta_0)^{cN-1}}{2^{cN-1}\beta_0\Gamma(cN/2)}
\end{equation}
and the most probable value of the fundamental inverse temperature is
\begin{equation}
\label{eq:betaf_most}
\beta_F^* = \beta_0\sqrt{2(cN-1)},
\end{equation}
going in the limit $N \rightarrow \infty$ towards $\beta^*$ in (\ref{eq:bstar}). Using the Gaussian approximation of (\ref{eq:pbetaf})
for large $N$ we obtain
\begin{equation}
\label{eq:bvar}
(\Delta \beta_F)^2 = (\beta_0)^2 = \frac{(\beta^*)^2}{2(cN-1)} \approx \frac{(\beta^*)^2}{2cN},
\end{equation}
a complete analog of (\ref{eq:var}). On the other hand, the probability density of $\beta_\Omega$ can be computed as
\begin{widetext}
\begin{equation}
P(\beta_\Omega = \beta|S) = \frac{2(cN-1)^{cN}}{\beta_0\Gamma(\tfrac{cN}{2})}
\exp\left(-({\scriptstyle cN-1})^2\Big(\frac{\beta_0}{\beta}\Big)^2\right)\left(\frac{\beta_0}{\beta}\right)^{cN+1}
\end{equation}
\end{widetext}
with its most probable microcanonical inverse temperature given by
\[\beta_\Omega^* = \beta_0(cN-1)\sqrt{\frac{2}{cN+1}} = \beta_F^*\sqrt{\frac{cN-1}{cN+1}},\] matching the most probable value of $\beta_F$ in
(\ref{eq:betaf_most}) in the limit $N \rightarrow \infty$. The variance of $\beta_\Omega$ in the Gaussian approximation is
\[(\Delta \beta_\Omega)^2 = (\beta_0)^2\left(\frac{cN-1}{cN+1}\right)^2 = (\beta_0)^2\left(\frac{\beta_\Omega^*}{\beta_F^*}\right)^2,\]
that is,
\begin{equation}
\frac{(\Delta \beta_\Omega)^2}{(\beta_\Omega^*)^2} = \frac{(\Delta \beta_F)^2}{(\beta_F^*)^2}.
\end{equation}

For large enough $N$ we can describe the system using symmetric confidence intervals. For instance, with 95.45\% confidence (2$\sigma$), using
(\ref{eq:eperstar}), (\ref{eq:var}), (\ref{eq:bstar}) and (\ref{eq:bvar}) we can write
\begin{subequations}
\begin{align}
\varepsilon & = \frac{1}{\beta_e}\sqrt{\frac{c c'}{\gamma}}\left(1 \pm \sqrt{\frac{2}{cN}}\right), \\
\beta & = \beta_e\sqrt{\frac{c\gamma}{c'}}\left(1 \pm \sqrt{\frac{2}{cN}}\right).
\end{align}
\end{subequations}

\begin{figure}[b!]
\begin{center}
\includegraphics[width=7.5cm]{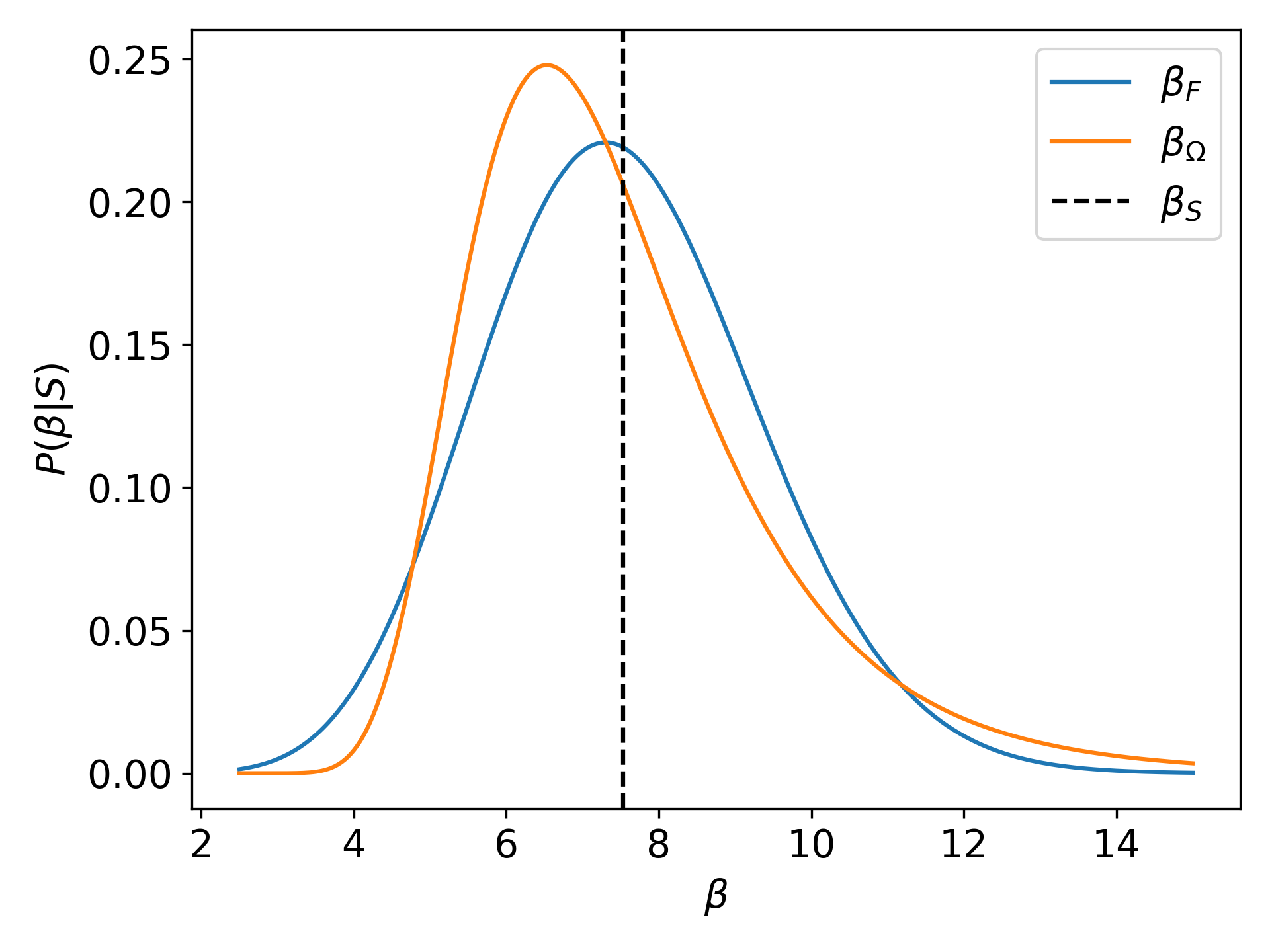}
\end{center}
\caption{Probability density for $\beta_F$ and $\beta_\Omega$ with $a = 1$, $\gamma = 1/10$, $c = 3$ and $\varepsilon_0 = 1$, for $N = 3$.}
\label{fig:betadist1}
\end{figure}

\begin{figure}
\begin{center}
\includegraphics[width=7.5cm]{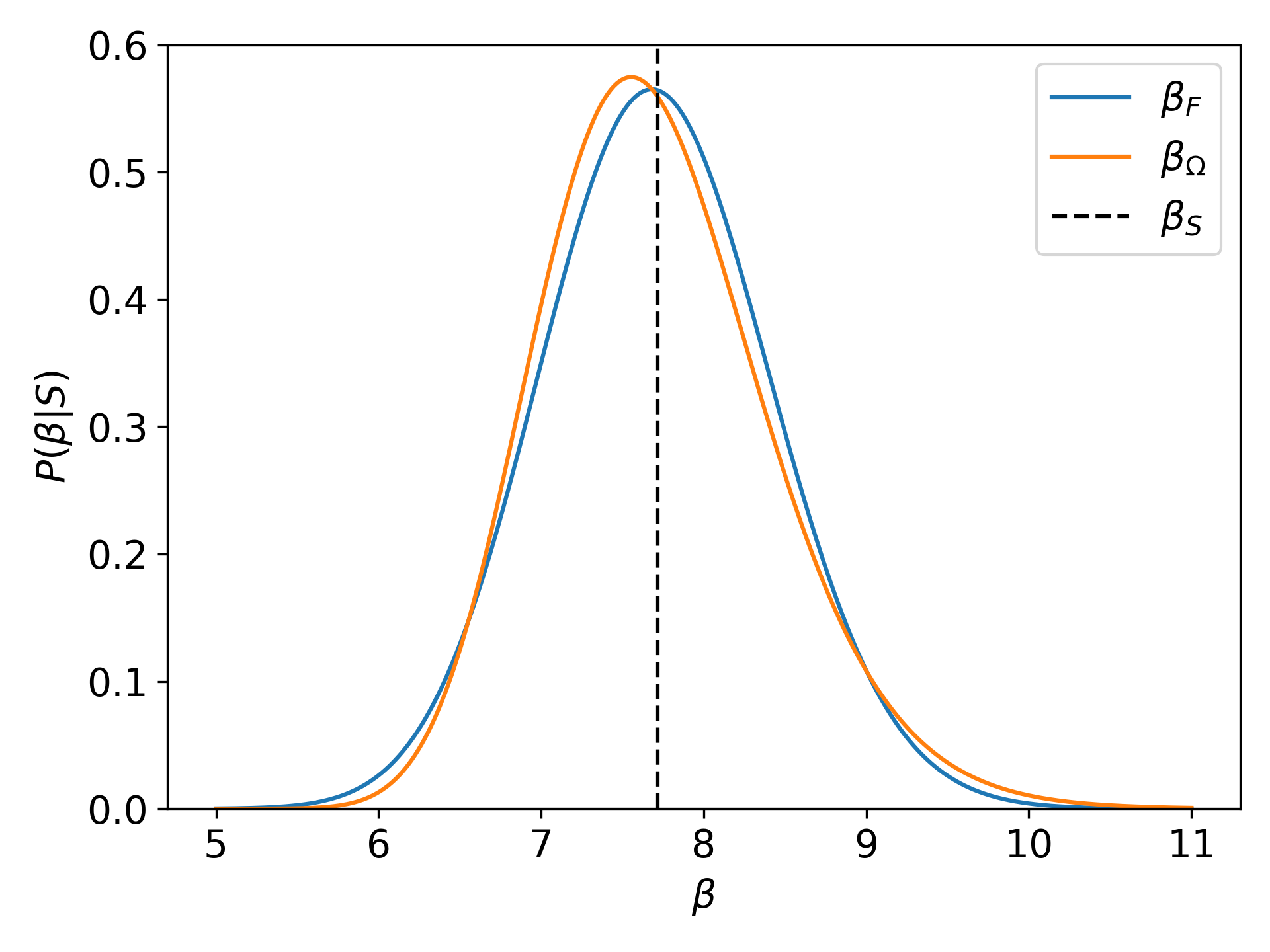}
\end{center}
\caption{Probability density for $\beta_F$ and $\beta_\Omega$ with $a = 1$, $\gamma = 1/10$, $c = 3$ and $\varepsilon_0 = 1$, for $N = 20$.}
\label{fig:betadist2}
\end{figure}

In this way, for a target system with $c = 3$ and $N$=267 we have an uncertainty of 5\% in both energy and inverse temperature, while for $N$=6667 we reduce
the uncertainty to 1\%.

\section{Deviations from canonical and microcanonical thermodynamics}

\begin{figure}
\begin{center}
\includegraphics[width=8cm]{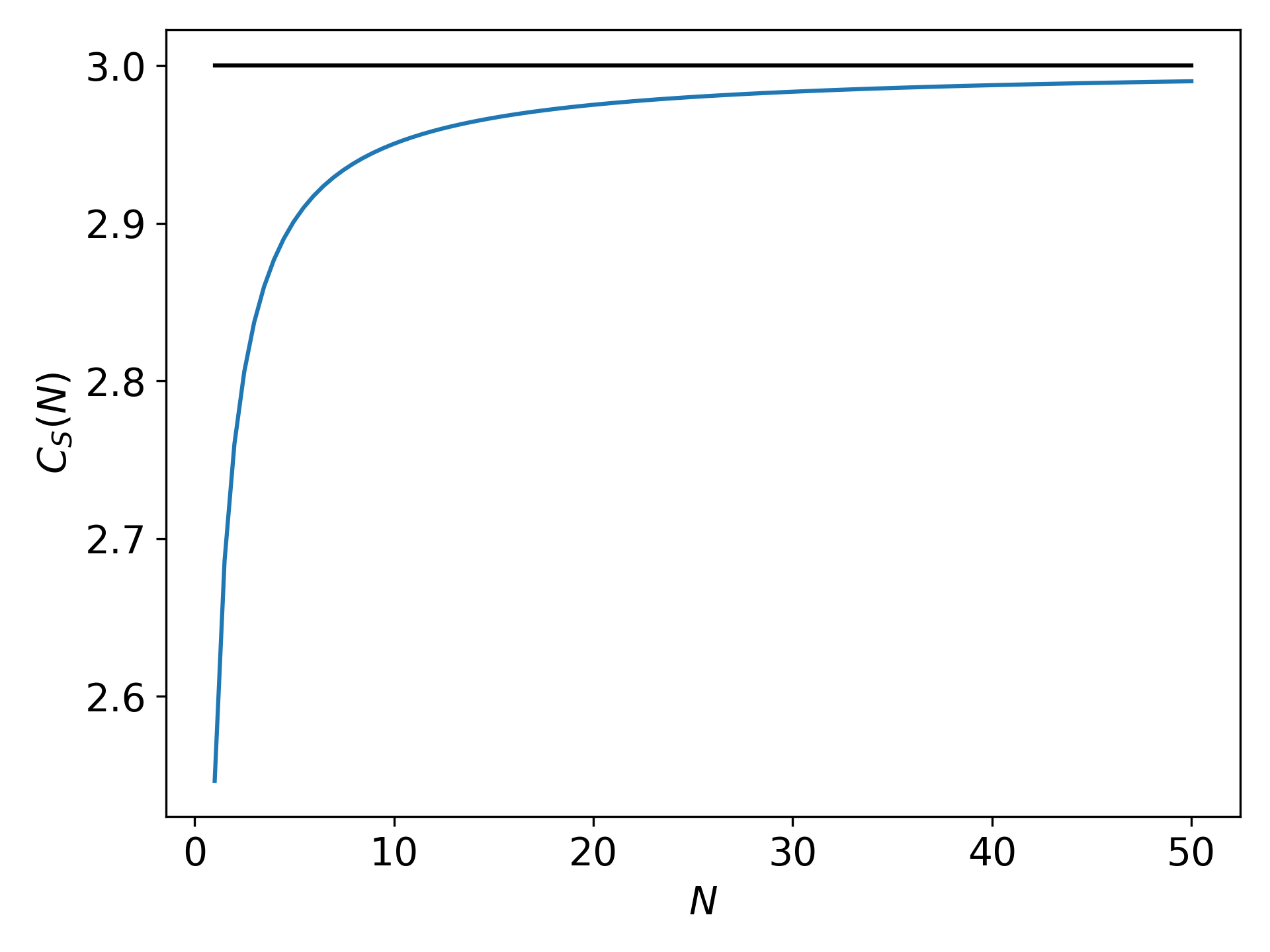}
\end{center}
\caption{Size-dependent ensemble heat capacity per particle $C_S(N)/N$ as defined in (\ref{eq:heatcap}) for $c = 3$.}
\label{fig:heatcap}
\end{figure}

We can construct the caloric curve by writing $\big<E\big>_S$ as a function of $\beta_S$, and we obtain
\begin{equation}
\begin{split}
\big<E\big>_S = \left[\frac{g_1 g_{-1}(cN-1)}{{g_0}^2}\right]& \cdot k_B T_S \\
    & = 2\left(\frac{g_1}{g_0}\right)^2 k_B T_S
\end{split}
\end{equation}
and this allows us to define an ensemble heat capacity
\begin{equation}
\label{eq:heatcap}
C_S(N) \defeq \frac{\partial \big<E\big>_S}{\partial T_S} = 2\left(\frac{g_1}{g_0}\right)^2 k_B
\end{equation}
which only depends on $c$ and $N$, and is such that \[\lim_{N \rightarrow \infty} \frac{C_S(N)}{N} = c\: k_B,\] thus recovering the heat capacity of the
isolated target in the thermodynamic limit, as shown in Fig.~\ref{fig:heatcap}.

The entropy of the target system, defined by (\ref{eq:entropy}) is, using (\ref{eq:rho_target}), directly given by
\begin{equation}
\begin{split}
\mathcal{S}_1 & = k_B\ln \mathcal{Z} + k_B\big<(\beta_0 E)^2\big>_S \\
              & = \mathcal{S}_0(N) + cNk_B\ln \big<E\big>_S
\end{split}
\end{equation}
where \[\mathcal{S}_0(N) \defeq \frac{cNk_B}{2}+k_B\ln \left[\frac{\omega(N)g_0}{2}\Big(\frac{g_0}{g_1}\Big)^{cN}\right]\]
is a constant contribution. This entropy is extensive, as is the case for the microcanonical entropy $k_B\ln \Omega_H(E)$, and is given by
\begin{equation}
\mathcal{S}_1 = Nk_B\left\{1+\ln \zeta+c\left[\frac{1}{2} + \ln \Big(\frac{\varepsilon^*}{c}\Big)\right]\right\},
\end{equation}
in fact smaller by $cN/2$ than the value in (\ref{eq:micro_entropy}) for $\varepsilon = \varepsilon^*$. The entropy of the composite system is also
extensive, and given by
\begin{equation}
\begin{split}
\mathcal{S}_{12} & = k_B\ln \Omega_\mathcal{H}(E_0) \\
                 & = \mathcal{S}_1 + N_e k_B\Big(s_G+\frac{c'}{2}\Big)
\end{split}
\end{equation}
where the term $N_e(s_G+c'/2)$ with $s_G$ given by (\ref{eq:sg}) is the contribution from the entropy of the environment at $G = E_0$.

An interesting feature of this ensemble is the analog of the fluctuation-dissipation formula \[\beta^2\big<(\delta E)^2\big>_\beta = \frac{C_V}{k_B}\] for the
canonical ensemble, which can be obtained from (\ref{eq:relvar}),
(\ref{eq:betas}) and (\ref{eq:heatcap}) as
\begin{equation}
(\beta_S)^2\big<(\delta E)^2\big>_S = cN\left(\frac{C_S(N)}{k_B}\right) - \left(\frac{C_S(N)}{k_B}\right)^2,
\end{equation}
and is such that in the limit $N \rightarrow \infty$ the right-hand side grows as $cN/2$. This is consistent with the variance $(\Delta \varepsilon)^2$
of the energy per particle in the Gaussian approximation according to (\ref{eq:var}), that we can write as \[(\beta^*)^2(\Delta \varepsilon)^2 = \frac{c}{2N}.\]

\section{Concluding remarks}

We have presented an approximation to the thermal statistics of an isolated system, conceptually divided into a target and an environment. We have shown that
when the environment has positive microcanonical heat capacity, the induced ensemble on the target cannot be described using superstatistics, but can have
well-defined distributions of energy and temperature. Interestingly, although the fluctuations in both energy per particle and inverse temperature decrease
with system size $N$ as $1/\sqrt{N}$, there are some signatures different from the canonical ensemble in the thermodynamic limit, such as the connection between
the variance of energy and the heat capacity, and also the value of the target entropy per particle.

\section*{Acknowledgments}

SD gratefully acknowledges support from ANID PIA ACT172101 and ANID FONDECYT 1171127.


\end{document}